\documentclass[12pt]{article}
\usepackage[a4paper, total={7in, 10in}]{geometry}
\usepackage[parfill]{parskip}
\usepackage{physics, tensor, float, subcaption}
\usepackage{graphicx}
\graphicspath{ {Plots/} }
\usepackage{jhep-mod}
\usepackage{bm}
\usepackage{soul}
\usepackage{amssymb,amsmath,amsthm}
\usepackage{mathrsfs}
\usepackage[utf8]{inputenc}
\usepackage{enumerate}
\usepackage{bigints}
\usepackage{xcolor}
\usepackage{appendix}
\usepackage{graphicx}
\usepackage{float}
\usepackage{tikz}
\usepackage{setspace}
\usepackage{cancel}
\usepackage{doi}
\definecolor{purple}{rgb}{1,0,1}
\definecolor{lime}{HTML}{A6CE39} 

\definecolor{lime}{HTML}{A6CE39}
\newcommand{\orcidicon}{%
	\begin{tikzpicture}
	\draw[lime, fill=lime] (0,0) 
		circle [radius=0.16] 
		node[white] {{\fontfamily{qag}\selectfont \tiny ID}};
	\draw[white, fill=white] (-0.0625,0.095) 
		circle [radius=0.007];
	\end{tikzpicture}
	\hspace{-5mm}
}

\newcommand\orcidAlex{{\href{https://orcid.org/0000-0002-1763-3563}{\orcidicon}}}
\newcommand\orcidMatt{{\href{https://orcid.org/0000-0003-1088-6485}{\orcidicon}}}

\renewcommand{\O}{\mathcal{O}}
\begin{document}

\title{\Huge
General-relativistic thin-shell\\ 
Dyson mega-spheres
}
\author{
\Large
Thomas Berry\!\orcidAlex\!$^1$, Alex Simpson\!\orcidAlex\!$^2$, 
Matt Visser\!\orcidMatt\!$^2$}
\affiliation{
$^1$ Robinson Institute,
Victoria University of Wellington, 
\\
\null\qquad PO Box 600, Wellington 6140, New Zealand.}
\affiliation{
$^2$ School of Mathematics and Statistics, 
Victoria University of Wellington, \\
\null\qquad PO Box 600, Wellington 6140, New Zealand.}
\emailAdd{thomas.berry@sms.vuw.ac.nz}
\emailAdd{alex.simpson@sms.vuw.ac.nz}
\emailAdd{matt.visser@sms.vuw.ac.nz}

\abstract{
\vspace{1em}

Loosely inspired by the somewhat fanciful notion of detecting an arbitrarily advanced alien civilization, we consider a general-relativistic thin-shell Dyson mega-sphere completely enclosing a central star-like object,  and perform a full general-relativistic analysis using the Israel--Lanczos--Sen junction conditions. We focus attention on the surface mass density, the surface stress, the  classical energy conditions, and the quasi-local forces between hemispheres.  We find that in the physically acceptable region the NEC, WEC, and SEC are always satisfied, while the DEC can be violated if the Dyson mega-sphere is sufficiently close to forming a black hole. 
We also demonstrate that  the original quasi-local version of the maximum force conjecture, 
$F \leq {1\over 4} F_{Stoney}= {1\over 4} F_{Planck}$, can easily be violated if the Dyson mega-sphere is sufficiently compact, that is, sufficiently close to forming a black hole. 
Interestingly there is a finite region of parameter space where one can violate the
original quasi-local version of the maximum force conjecture \emph{without} violating the DEC.  
Finally, we very briefly discuss the possibility of nested thin-shell mega-spheres (Matrioshka configurations) and thick-shell 
Dyson mega-spheres.

\bigskip
\bigskip

\bigskip
\noindent
{\sc Date:} 6 July 2022;  16 September 2022; \LaTeX-ed \today.

\bigskip
\noindent{\sc Keywords}:
Dyson mega-sphere; Kardashev type II civilization; general relativity; Israel--Lanczos--Sen junction conditions; classical energy conditions; Quasi-local maximum force conjecture; Matrioshka  configurations; Matrioshka brain.

\bigskip
\noindent{\sc PhySH:} 
Gravitation

\bigskip
This version accepted for publication in Physical Review D.
}

\maketitle
\def\tr{{\mathrm{tr}}}
\def\diag{{\mathrm{diag}}}
\def\cof{{\mathrm{cof}}}
\def\pdet{{\mathrm{pdet}}}
\def\d{{\mathrm{d}}}
\parindent0pt
\parskip7pt
\def\Kerr{{\scriptscriptstyle{\mathrm{Kerr}}}}
\def\eos{{\scriptscriptstyle{\mathrm{eos}}}}
\newcommand*{\Chi}{\mbox{\Large$\chi$}}

\clearpage
\section{Introduction}

In 1960 Freeman Dyson~\cite{Dyson} mooted the idea that an arbitrarily advanced civilization, 
(at least a  Kardashev type II civilization~\cite{Kardashev:1,Kardashev:2}), 
might seek to control and utilize the energy output of an entire star by building a spherical mega-structure to completely enclose the star, trap all its  radiant emissions, and use the energy flux to do ``useful'' work.
Dyson's idea has led to observational searches~\cite{Carrigan:2008}, extensive technical discussions~\cite{BH-Dyson}, and more radical proposals such as reverse-Dyson configurations~\cite{Reverse-Dyson}, (where one harvests the CMB and dumps waste heat into a central black hole),  and ``hairy'' Dyson spheres~\cite{Galileon-Dyson}, (with Gallileon ``hair''). 
The vast majority of currently available analyses along these lines use Newtonian  gravity, but herein we shall seek to perform an analysis in the thin-shell (Israel--Lanczos--Sen~\cite{Israel:1966, Barrabes:1991, Lanczos:1924, Lanczos:1922, Sen} junction condition) limit using standard general relativity.\footnote{Much more rarely one may sometimes encounter analyses using nonstandard theories of modified  gravity~\cite{Galileon-Dyson}.}
For the current article, in the interests of tractability,  we shall restrict attention to investigating systems with spherical symmetry.

For calculations we adopt geometrodynamic units where $G_N\to1$ and $c\to 1$, see for instance~\cite{MTW,Wald}.
Sometimes we will present key results in physical SI units, but when we do we shall explicitly say so.
We furthermore adopt the sign conventions of~\cite{MTW,Wald}.

\section{Spacetime metric ansatz}\label{ansatz}

To set the stage, let the central star have mass $m$, and take the mega-sphere to have a radius $a$ larger than the radius of the star; in fact for almost all purposes we may safely idealize the radius of the central star to be zero. Let the total mass of the (star)+(mega-sphere) system be $M$. Then, outside the mega-sphere, the spacetime geometry can be described by 
\begin{equation}
\d s^2 = -\left(1-{2M\over r}\right) \d t^2 + {\d r^2\over 1-2M/r} + r^2 \;\d\Omega^2; 
\qquad (r>a);
\end{equation}
while inside the mega-sphere
\begin{equation}
\d s^2 = -\xi^2 \left(1-{2m\over r}\right) \d t^2 + {\d r^2\over 1-2m/r} + r^2 \;\d\Omega^2.
\qquad (r<a);
\end{equation}

\clearpage
Here to maintain continuity of the $t$ coordinate as one crosses the mega-sphere at $r=a$ it is useful to set
\begin{equation}
\xi^2 = {1-2M/a\over 1-2m/a}.
\end{equation}
Physically we must demand $ 0 \leq m \leq M < a/2$. First we demand $m < a/2$ and $M < a/2$ to prevent horizon formation from enclosing the mega-sphere. Then we demand $M\geq m$ so that the total mass is not smaller than the mass of the central object. Finally $m\geq 0$ so that the central object does not have negative mass. Putting this all together we have $ 0 \leq m \leq M < a/2$, so that $0\leq \xi^2 \leq 1$.

It is furthermore useful to define a proper distance radial coordinate, normalized so that $\ell=0$ on the shell at $r=a$: 
\begin{equation}
\label{E:proper}
\ell(r) = \left\{ \begin{array}{l l}
\displaystyle{+\int_a^r {\d r \over \sqrt{1-2M/r}} }& \qquad (r>a);\\[20pt]
\displaystyle{-\int_r^a {\d r \over \sqrt{1-2m/r}}} & \qquad (r<a).
\end{array}
\right.
\end{equation}
Then $n_a = \nabla_a \ell$ is by construction a unit outward-pointing spacelike vector,
which we can use to define the projection operator
\begin{equation}
h_{ab} = g_{ab} - n_a\; n_b.
\end{equation}

Finally, formally inverting $\ell(r)$ to implicitly extract $r(\ell)$, one can re-express the spacetime geometry as 
\begin{equation}
\d s^2 = -\left(1-{2M\over r(\ell)}\right) \d t^2 + {\d\ell^2} 
+ r(\ell)^2 \;\d\Omega^2; 
\qquad (r>a);
\end{equation}
\begin{equation}
\d s^2 = -\xi^2 \left(1-{2m\over r(\ell)}\right) \d t^2 + {\d\ell^2} 
+ r(\ell)^2 \;\d\Omega^2; 
\qquad (r<a).
\end{equation}

\section{Israel--Lanczos--Sen junction conditions}
It is now straightforward to calculate the extrinsic curvature tensors
\begin{equation}
(K^\pm) _{ab} = 
{1\over2} \left. {\partial g_{ab}\over\partial \ell} 
\right|_{\ell= 0^\pm}
=
{1\over2} \left. {\partial r \over\partial\ell} \; {\partial g_{ab}\over\partial r } 
\right|_{\ell= 0^\pm}
=
{1\over2} \left. {1 \over\sqrt{g_{rr}}} \; {\partial g_{ab}\over\partial r } 
\right|_{\ell= 0^\pm}
\end{equation}
 just above and below the thin shell at $r=a$. (This calculation is well-known and can be found in very many places. See for instance references~\cite{Israel:1966, Barrabes:1991, Lanczos:1924, Lanczos:1922, Sen}  and the more recent references~\cite{Lorentzian-wormholes, Visser:1989a, Visser:1989b, Poisson:1995, Lobo:2015-MGxx}.) 

\clearpage
Then in the coordinate basis the extrinsic curvatures are: 
\begin{equation}
(K^+)_{tt} = -\sqrt{1-2M/a} \; {M\over a^2}; \qquad 
(K^+)_{\theta \theta} = {(K^+)_{\phi \phi} \over \sin^2\theta} 
= \sqrt{1-2M/a} \; a;
\end{equation}
and
\begin{equation}
(K^-)_{tt} =  -\xi^2 \sqrt{1-2m/a}\; {m\over a^2}; \qquad 
(K^-)_{\theta \theta} = {(K^-)_{\phi\phi}\over \sin^2\theta} 
= \sqrt{1-2m/a}\;  a.
\end{equation}

Rewriting this in the natural orthonormal basis one has: 
\begin{equation}
(K^+)_{\hat t \hat t} = - {M/a^2\over\sqrt{1-2M/a}}; \qquad 
(K^+)_{\hat \theta \hat \theta} = (K^+)_{\hat \phi \hat \phi} 
= {\sqrt{1-2M/a}\over a};
\end{equation}
and
\begin{equation}
(K^-)_{\hat t \hat t} = - {m/a^2\over\sqrt{1-2m/a}}; \qquad 
(K^-)_{\hat \theta \hat \theta} = (K^-)_{\hat \phi \hat \phi} 
= {\sqrt{1-2m/a}\over a}.
\end{equation}
Furthermore, the (distributional) stress-energy on the shell is then easily obtained in terms of the discontinuity $[K]_{ab}$ of the extrinsic curvatures ~\cite{Israel:1966, Barrabes:1991, Lanczos:1924, Lanczos:1922, Sen, Lorentzian-wormholes, Visser:1989a, Visser:1989b, Poisson:1995, Lobo:2015-MGxx}
\begin{equation}
[K]_{ab} = (K_+)_{ab} - (K_-)_{ab}.
\end{equation}
Explicitly, working in the natural orthonormal basis,
\begin{equation}
[K]_{\hat a\hat b}  = \left[ \begin{array}{cc|cc}
 -{M/a^2\over\sqrt{1-2M/a}} +  {m/a^2\over\sqrt{1-2m/a}} &0\;&0&0\\[10pt]
0&0&0&0\\[1pt]
\hline
0&0&{\vphantom{\Bigg|}{\sqrt{1-2M/a}\over a}}-{\sqrt{1-2m/a}\over a}&0\\
0&0&0&{\sqrt{1-2M/a}\over a}-{\sqrt{1-2m/a}\over a}
\end{array}
\right].
\end{equation}

For the discontinuity in the trace of the extrinsic curvature, $K= g^{ab} K_{ab}=
 g^{\hat a\hat b} K_{\hat a\hat b}$,  we have
\begin{equation}
[K] = + {M/a^2\over\sqrt{1-2M/a}} -  {m/a^2\over\sqrt{1-2m/a}} +
2\left( {\sqrt{1-2M/a}\over a}-{\sqrt{1-2m/a}\over a} \right),
\end{equation}
which can be re-written as
\begin{equation}
[K] = +{1\over a} \left\{ {2-3M/a\over \sqrt{1-2M/a}} -  {2-3m/a\over \sqrt{1-2m/a}} \right\}.
\end{equation}

\clearpage
By invoking the 
Israel--Lanczos--Sen junction conditions, (\cite{Israel:1966,Barrabes:1991,Lanczos:1924,Lanczos:1922,Sen}, see also~\cite{Lorentzian-wormholes, Visser:1989a, Visser:1989b, Poisson:1995, Lobo:2015-MGxx}), 
we find the distributional stress-energy tensor:
\begin{equation}
T_{ab} = -{1\over 8\pi} \left( [K]_{ab}  -  [K] h_{ab} \right) \, \delta(\ell) 
= S_{ab}\; \delta(\ell),
\end{equation}
or better yet, its orthonormal form:
\begin{equation}
T_{\hat a\hat b} = -{1\over8\pi} \left( [K]_{\hat a\hat b}  -  [K] h_{\hat a\hat b} \right) \, \delta(\ell) 
= S_{\hat a\hat b}\; \delta(\ell).
\end{equation} 
Here $\ell$ is the proper radial distance from the shell, as defined in equation (\ref{E:proper}). Furthermore
we define the surface energy density $\sigma$, and surface stress  $\Pi$ (this is the \emph{negative} of what is usually called the surface tension $\vartheta$, that is, $\Pi=-\vartheta$)  by 
\begin{equation}
S_{\hat a\hat b}  = \left[ \begin{array}{cc|cc}
 \sigma &0\;&0&0\\
0&0&0&0\\
\hline
0&0&\Pi&0\\
0&0&0&\Pi
\end{array}
\right].
\end{equation}
Thence
\begin{equation}
\sigma = {1\over 4\pi \, a } \; \left( \sqrt{1-2m/a}-\sqrt{1-2M/a}\right),
\end{equation}
and
\begin{equation}
\Pi = {1\over 8\pi \, a } \left( {1-M/a\over\sqrt{1-2M/a}} -
{1-m/a\over\sqrt{1-2m/a}} 
\right).
\end{equation}

Now, as long as $M \geq m$, it is immediate that the surface energy density is positive, $\sigma \geq 0$.
To bound the surface stress $\Pi$, we note that
\begin{equation}
{\partial\Pi\over\partial M} = {M\over 8\pi \, a^3 (1-2M/a)^{3/2}} >0.
\end{equation}
Thence, as long as we satisfy the physically motivated constraint that $M \geq m$, (the total mass of the system is greater than the mass of the central star), it is immediate that $\Pi \geq 0$.
So in the physically relevant regime, where $a > 2M \geq 2m\geq 0$,  all the surface stress-energy components are guaranteed non-negative. 

Somewhat similar calculations are relevant to thin-shell gravastar models~\cite{gravastar1,gravastar2,gravastar3,gravastar4}, though in that context many of the technical details, and all of the physics goals, are very different.

\clearpage
\section{Classical energy conditions}
\enlargethispage{40pt}
\subsection{Definitions}

For any type I stress energy tensor~\cite{Hawking},
(which is certainly and explicitly the case in the current spherically symmetric situation), the standard classical energy conditions are based on considering various linear combinations of the energy density $\rho$ and the principal pressures $p_i$. (See primarily references~\cite{Barcelo:2002, Curiel:2014, Kontou:2020, Helfer:1998, Alcubierre:2017, Lobo:2004}, but see also references~\cite{Visser:1997a, Visser:1997b, Visser:cosmo99, Visser:2003-Kar} and~\cite{Martin-Moruno:2013a, Martin-Moruno:2013b, Martin-Moruno:2015,
Martin-Moruno:2017a, Martin-Moruno:2017b,Martin-Moruno:2021, Santiago:2021a,  Santiago:2021b}.)

\begin{description}
\item[NEC:]  $\rho+p_i\geq 0$.
\item[WEC:] $\rho\geq 0$ and $\rho+p_i\geq 0$ .
\item[SEC:]  $\rho+ \sum_i p_i \geq 0$ and $\rho+p_i\geq 0$.
\item[DEC:] $|p_i| \leq \rho$.
\end{description}
Thence, in this specific thin-shell configuration, with principal stresses of the form $\mathrm{diag}\{\sigma,0,\Pi,\Pi\}$, these standard classical energy conditions will specialize to:
\begin{description}
\item[NEC:]  $\sigma\geq 0$ and $\sigma+\Pi \geq 0$.
\item[WEC:] Equivalent to NEC.
\item[SEC:] NEC and $\sigma+2\Pi \geq 0$.
\item[DEC:] NEC and $\sigma-\Pi \geq 0$. (That is, $|\Pi| \leq \sigma$.) 
\end{description}
We shall now analyze these energy conditions in detail.

\subsection{Null, weak, and strong energy conditions}

In the physically relevant regime, $M \geq m$,  both $\sigma$ and $\Pi$  are guaranteed non-negative, so all three of the NEC, WEC, and SEC are automatically satisfied. 

\subsection{Dominant energy condition}

To investigate the DEC it is useful to define the dimensionless quantity
\begin{equation}
w = {\Pi\over\sigma},
\end{equation}
and ask whether or not $w\leq1$. (We already know $w\geq 0$.)

Explicitly, to investigate the DEC we see that we need to consider the dimensionless parameter
\begin{equation}
w = { 1\over 2}{ 
\left( {1-M/a\over\sqrt{1-2M/a}} -{1-m/a\over\sqrt{1-2m/a}} \right)
\over
\left( \sqrt{1-2m/a}-\sqrt{1-2M/a}\right)
}
\end{equation}

\clearpage

This quantity is more easily analyzed in terms of two dimensionless ``compactness'' parameters $\Chi = 2M/a$ and $\chi=2m/a$, where now the physically acceptable range corresponds to $0 \leq \chi \leq \Chi \leq 1$. Then,
\begin{equation}
w = { 1\over 2}{ 
\left( {1-\Chi/2\over\sqrt{1-\Chi}} -{1-\chi/2\over\sqrt{1-\chi}} \right)
\over
\left( \sqrt{1-\chi}-\sqrt{1-\Chi}\right)
}
= {1\over4} \left\{{1\over\sqrt{1-\chi}\sqrt{1-\Chi} } -1 \right\}.
\end{equation}
Saturation of the DEC, the condition $w=1$, then easily translates to
\begin{equation}
(1-\Chi)(1-\chi) = {1\over25}.
\end{equation}
Less symmetrically, but perhaps more usefully, this can be rewritten as
\begin{equation}
\chi = 1 - {1\over25(1-\Chi)}; \qquad\qquad
\Chi = 1 - {1\over25(1-\chi)}.
\end{equation}
For a thin-shell Dyson mega-sphere on the verge of violating the DEC we have
\begin{equation}
\sigma = \Pi = {1\over 4\pi \, a } \; 
\left( \sqrt{1-\chi}-\sqrt{1-\Chi}\right)=
{1\over 4\pi \, a } \; 
\left( {1\over5\sqrt{1-\Chi}} -\sqrt{1-\Chi}\right).
\end{equation}
To enforce $\sigma\geq 0$ while saturating the DEC one needs $\Chi\geq 4/5$, (and $\Chi<1$). 

Furthermore, the DEC-violating region is easily seen to be
\begin{equation}
1 - {1\over25(1-\Chi)} < \chi<\Chi \leq 1.
\end{equation}
That is
\begin{equation}
{{24\over25}-\Chi\over1-\Chi} < \chi<\Chi \leq 1.
\end{equation}

Thus violations of the DEC are relatively easy to arrange. (At least theoretically; the implied construction materials would be somewhat \emph{outr\'e}.) 
It should be emphasized that violations of the classical energy conditions are not absolute prohibitions on ``interesting physics'',  see~\cite{Barcelo:2002, Curiel:2014, Kontou:2020} and specifically~\cite{Santiago:2021a,  Santiago:2021b, Visser:2021c}, but they certainly are invitations to think very carefully about the underlying physics. 

\begin{figure}[!htbp]
\begin{center}
\includegraphics[scale=0.5]{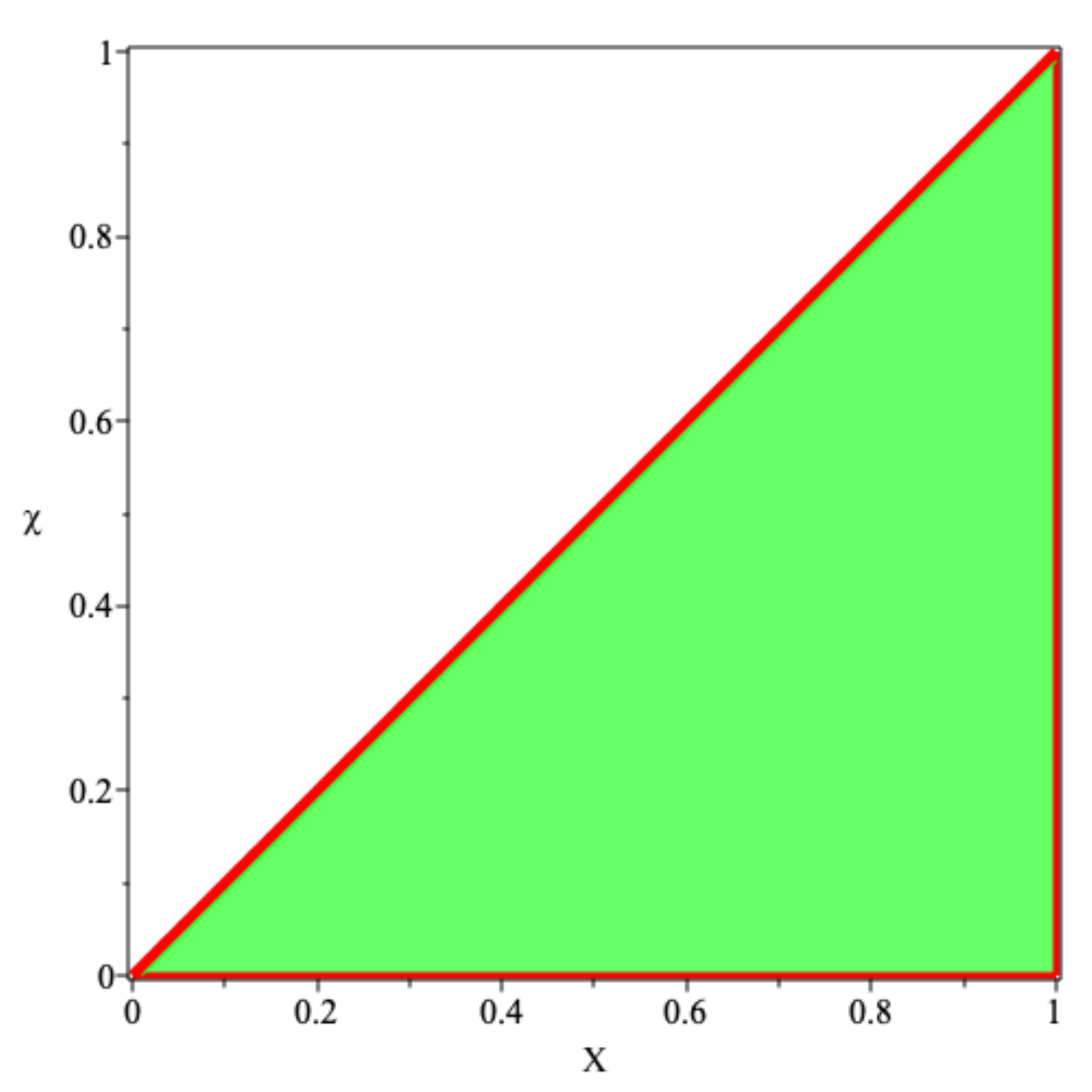}
\caption{Boundary and interior of the physically acceptable region $0\leq\chi\leq\Chi<1$.}
\label{F-DEC-0}
\end{center}
\end{figure}

\begin{figure}[htbp]
\begin{center}
\includegraphics[scale=0.5]{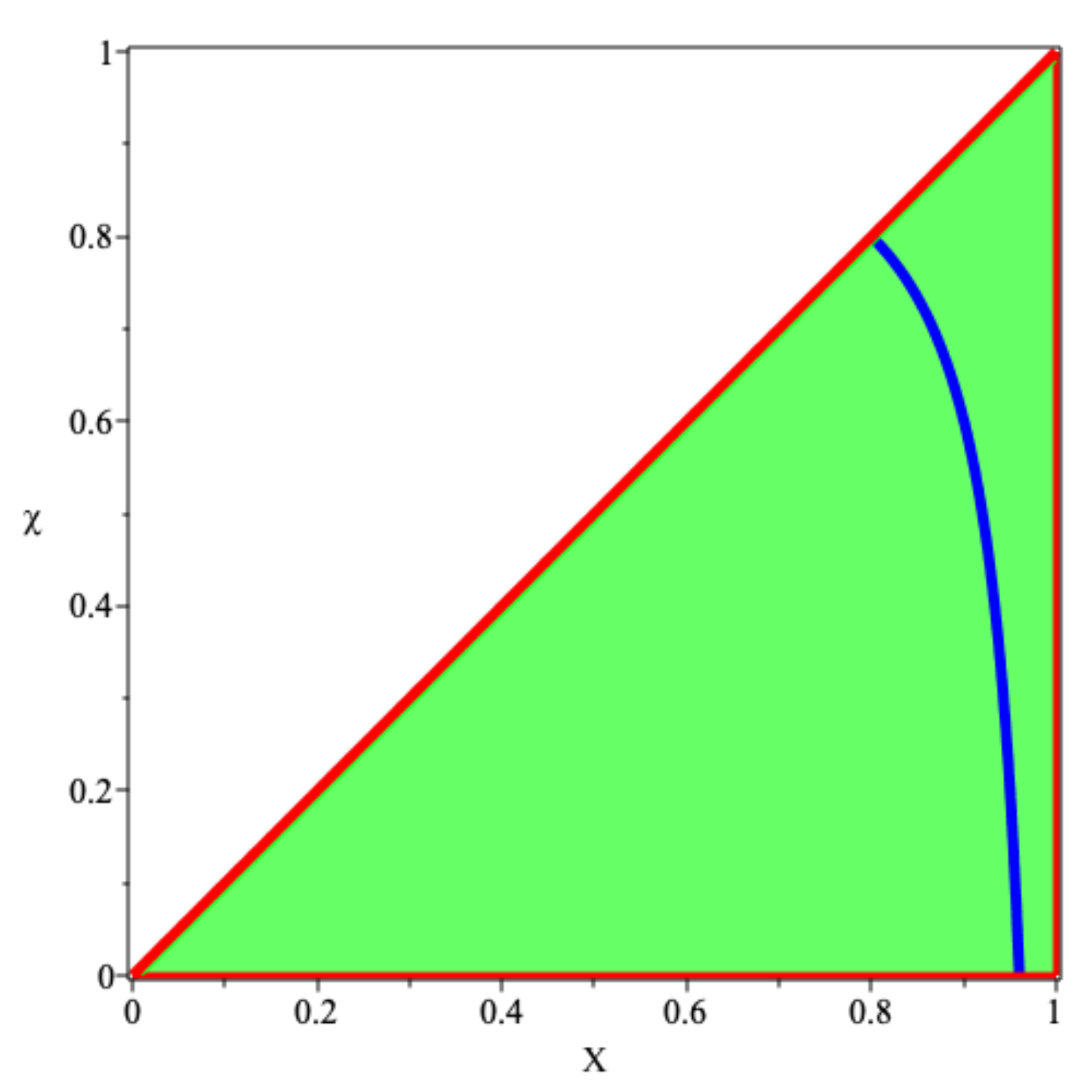}
\caption{DEC violations:
The green region is the physically acceptable region, that is  $0\leq\chi\leq\Chi<1$. The blue curve corresponds to $w=1$, the boundary of the DEC violating region. 
To the left of the blue curve $w<1$ and the DEC is satisfied. 
To the right of the blue curve $w>1$ and the DEC is violated.
}
\label{F-DEC-1}
\end{center}
\end{figure}

\begin{figure}[htbp]
\begin{center}
\includegraphics[scale=0.5]{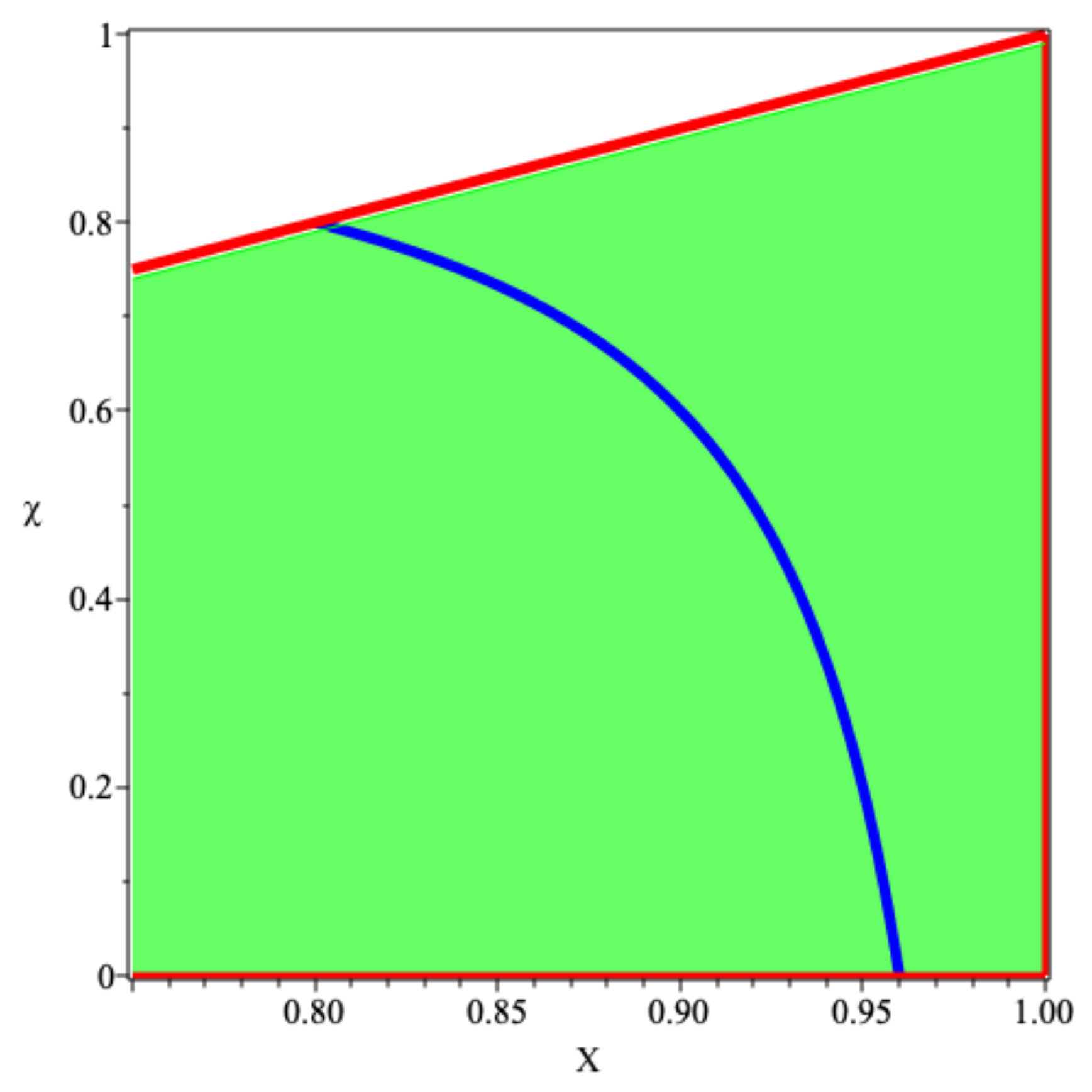}
\caption{DEC violations, now focussing on the region above  $\Chi = {3\over 4}$.
The green region is the physically acceptable region, that is  $0\leq\chi\leq\Chi<1$. The blue curve corresponds to $w=1$, the boundary of the DEC violating region. 
To the left of the blue curve $w<1$ and the DEC is satisfied. 
To the right of the blue curve $w>1$ and the DEC is violated.
}
\label{F-DEC-2}
\end{center}
\end{figure}

The three regions where $w<1$, $w=1$, and $w>1$ in the physically acceptable part of the $(\chi,\Chi)$ plane, $0\leq\chi\leq\Chi\leq1$, are plotted in figures \ref{F-DEC-0}--\ref{F-DEC-2}.

\clearpage
\section{Quasi-local forces}

Whenever one is dealing with an extended object in general relativity, (star, planet, Dyson mega-sphere), the notion of ``force'' is at best a quasi-local concept associated with some region and its boundary. For a static bulk region with boundary 2-surface $\Sigma$ it is natural to use the induced 2-metric $(g_2)_{ij}$, and the normal $n^a$ to $\Sigma$, to define 
\begin{equation}
F_\Sigma = \int_\Sigma (T_{ab} \, n^a n^b) \,\sqrt{g_2}\, \d^2 x =
\int_\Sigma p_\perp \; \d A.
\end{equation}
For a static thin-shell region (a 2-surface) with boundary 1-surface $L$ it is natural to use the induced 1-metric $(g_1)$ on $L$, and the within-shell normal $n^a$ to the curve $L$, to define 
\begin{equation}
F_L = \int_L (T_{ab} \, n^a n^b) \,\sqrt{g_1}\, \d x = 
 \int_L \Pi_\perp \; \d L.
\end{equation}
These quasi-local forces were the impetus for original quasi-local versions of the maximum force conjecture, now thoroughly disproved by a number of explicit counter-examples~\cite{Jowsey:2021a}. 
It is only for point-like objects that any ``ultra-local'' notion of force can meaningfully be formulated,  and in the current context (working with Dyson mega-spheres) it is clear that the quasi-local notion of force is primary.

\subsection{Quasi-local force between two hemispheres}

The quasi-local force between any two hemispheres separated by a great circle on the Dyson mega-sphere is simply given (in the usual geometrodynamic units) by the dimensionless quantity
\begin{equation}
F = \Pi \times (2\pi a) = {1\over 4 } \left( {1-M/a\over\sqrt{1-2M/a}} -
{1-m/a\over\sqrt{1-2m/a}} 
\right).
\end{equation}
We first rewrite this in terms of the two dimensionless compactness parameters, $\Chi = 2M/a$ and $\chi=2m/a$, as 
\begin{equation}
F = {1\over 4 } \left( {1-\Chi/2\over\sqrt{1-\Chi}} -
{1-\chi/2\over\sqrt{1-\chi}} 
\right).
\end{equation}
This can further be recast as 
\begin{equation}
F = {1\over 8 } \left( {1\over\sqrt{1-\Chi}}+ \sqrt{1-\Chi} -
{1\over\sqrt{1-\chi}} - \sqrt{1-\chi} 
\right).
\end{equation}
The maximum original quasi-local force conjecture, in its strong form, would require $F\leq 1/4$.
But this is easily seen to be generically violated; see discussion below. 

\subsection{Physical units}

To rephrase this discussion in physical units, let us consider the Stoney
force~\cite{Stoney1,Stoney2,Stoney3}, (also known as the Planck force, though the analysis by Stoney pre-dates that of Planck by some 20 years~\cite{Planck}).  This  is defined by 
\begin{equation}
 F_{Stoney} = F_{Planck} = {c^4 \over G_N} \approx 1.2\times10^{44} \hbox{ N},
 \label{stoney-planck}
\end{equation}
in terms of which, re-inserting the factors of $c$ and $G_N$ as in \eqref{stoney-planck}, one has
\begin{equation}
F_{physical} = {1\over 4 } \left( {1-\Chi/2\over\sqrt{1-\Chi}} -
{1-\chi/2\over\sqrt{1-\chi}} 
\right) F_{Stoney}.
\end{equation}
Alternatively,
\begin{equation}
F_{physical} = {1\over 8 } \left( {1\over\sqrt{1-\Chi}}+ \sqrt{1-\Chi} -
{1\over\sqrt{1-\chi}} - \sqrt{1-\chi} 
\right) F_{Stoney}.
\end{equation}

\subsection{Special case $m=0$}

We first note that
\begin{equation}
{\partial F\over\partial M} = {M\over 4 a^2 (1-2M/a)^{3/2} } >0;
\end{equation}
and
\begin{equation}
{\partial F\over\partial m} = -{m\over 4 a^2 (1-2m/a)^{3/2} } <0.
\end{equation}
So to maximize the quasi-local force between any two hemispheres one should maximize $M$, (without forming a black hole, so one requires $M<a/2$), and minimize $m$, (for instance by setting $m\to 0$, so that the Dyson mega-sphere is empty, there is no central star). 

It is then easy to check that for the special case $m=0$ we have
\begin{equation}
F_{m=0} = {1\over 4 } \left( {1-M/a\over\sqrt{1-2M/a}} - 1 \right)
= {1\over 4 } \left( {1-\Chi/2\over\sqrt{1-\Chi}} - 1 \right).
\end{equation}
Therefore, when forcing $m=0$,  we certainly have $F_{m=0} > 1/4$ over the entire range 
\begin{equation}
\Chi  \in \Big( \left\{4\sqrt3-6\right\}  , 1\Big) = \Big(0.928203232..., 1\Big).
\end{equation}
\clearpage 

Thence, in this Dyson mega-sphere setting, sufficiently close to black hole formation  the original quasi-local form of the maximum force conjecture is manifestly false. If one desires to rescue the the maximum force conjecture one needs to redefine the notion of force in an ``ultra-local'' manner, by carefully (and somewhat artificially) restricting the domain of discourse to only include certain ``ultra-local'' forces relevant to point-particle situations, and not relevant for extended objects. It is only for such a redefined ``ultra-local'' maximum force conjecture that there is any chance of getting it to work. 
See related discussion in references~\cite{Jowsey:2021a, Jowsey:2021b, Faraoni:2021}, with countervailing opinions in references~\cite{Schiller:2021a, Schiller:2021b}.

\subsection{General case}
\enlargethispage{20pt}

\begin{figure}[!htbp]
\begin{center}
\includegraphics[scale=0.5]{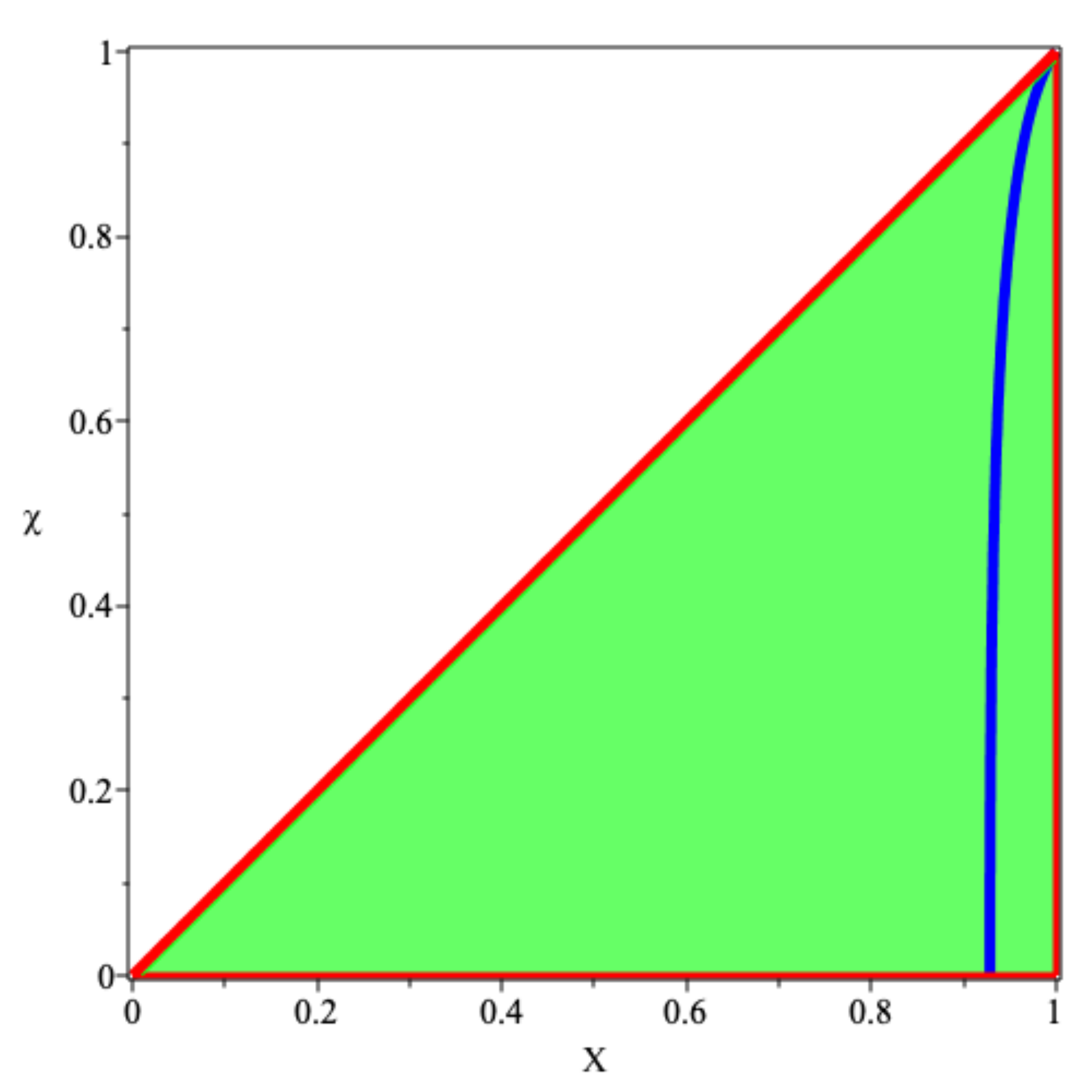}
\caption{
Violations of the (naive) quasi-local maximum force conjecture:
The green region is the physically acceptable region,  $0\leq\chi\leq\Chi<1$. The blue curve corresponds to $F=1/4$, the boundary of the region violating the (naive) quasi-local maximum force conjecture. This curve intersects the $\Chi$ axis at $\Chi = 4\sqrt{3}-6 = 0.928203232...$. 
To the left of the blue curve $F<1/4$ and the (naive) quasi-local  maximum force conjecture is satisfied. 
To the right of the blue curve $F>1/4$ and the (naive) quasi-local  maximum force conjecture is violated.
}
\label{F-force-1}
\end{center}
\end{figure}

\begin{figure}[!htbp]
\begin{center}
\includegraphics[scale=0.5]{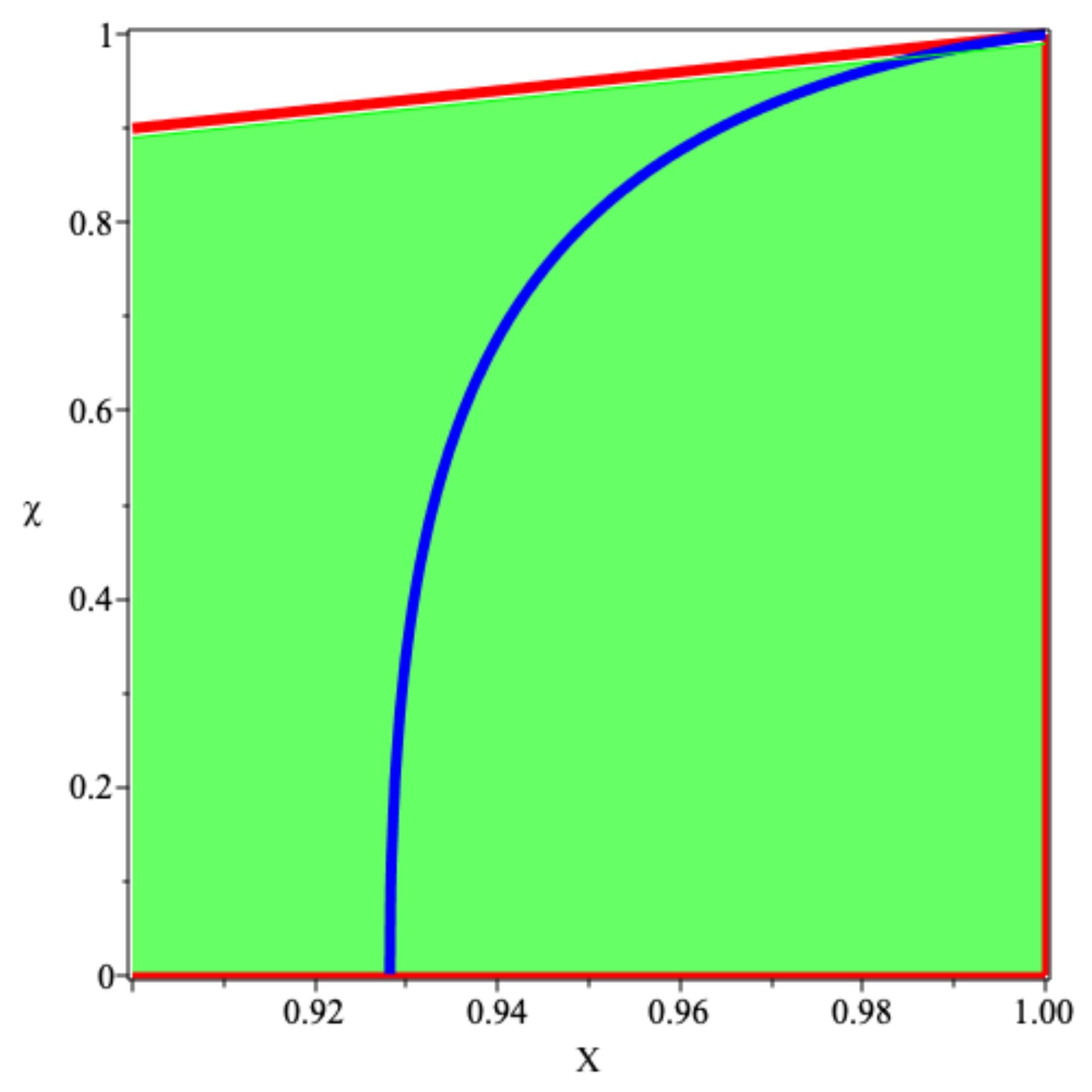}
\caption{
Violations of the (naive) maximum quasi-local  force conjecture, now focussing on the region $\Chi > {9\over10}$:
The green region is the physically acceptable region,  $0\leq\chi\leq\Chi<1$. The blue curve corresponds to $F=1/4$, the boundary of the region violating the (naive) maximum quasi-local  force conjecture. This curve intersects the $\Chi$ axis at $\Chi = 4\sqrt{3}-6 = 0.928203232...$. 
To the left of the blue curve $F<1/4$ and the (naive) maximum quasi-local  force conjecture is satisfied. 
To the right of the blue curve $F>1/4$ and the (naive) maximum quasi-local  force conjecture is violated.
}
\label{F-force-2}
\end{center}
\end{figure}

More generally in terms of the two compactness parameters $\Chi$ and $\chi$, we have already seen:
\begin{eqnarray}
F &=&  {1\over 4 } \left( {1-\Chi/2\over\sqrt{1-\Chi}} -
{1-\chi/2\over\sqrt{1-\chi}} \right)
\\
&=& {1\over 8 } \left( {1\over\sqrt{1-\Chi}}+ \sqrt{1-\Chi} -
{1\over\sqrt{1-\chi}} - \sqrt{1-\chi} 
\right).
\end{eqnarray}
For any fixed value of $\chi$, this force $F$ becomes arbitrarily large as $\Chi\to 1$.
Specifically, the boundary of the region satisfying both $F <1/4$ and the physicality constraints $0 \leq \chi \leq \Chi \leq 1$ is plotted in figures \ref{F-force-1} and \ref{F-force-2}.\footnote{
To obtain the curved line in figures \ref{F-force-1} and \ref{F-force-2} one sets $F=1/4$ and solves resulting quadratic for $\chi$ as a function of $\Chi$. This quadratic yields two somewhat messy roots, only one of which is physical. Even then, one has to be careful to restrict attention to the physically relevant region. 
}


Again, we see that the original quasi-local form of the maximum force conjecture is manifestly false.
For this quasi-local maximum force violation to occur one must be ``close'' to black hole formation $\chi\lesssim\Chi\lesssim1$.


\clearpage

\section[Violating the quasi-local maximum force conjecture\\
without violating the DEC]
{Violating the quasi-local maximum force conjecture --- without violating the DEC}

\begin{figure}[!htbp]
\begin{center}
\includegraphics[scale=0.5]{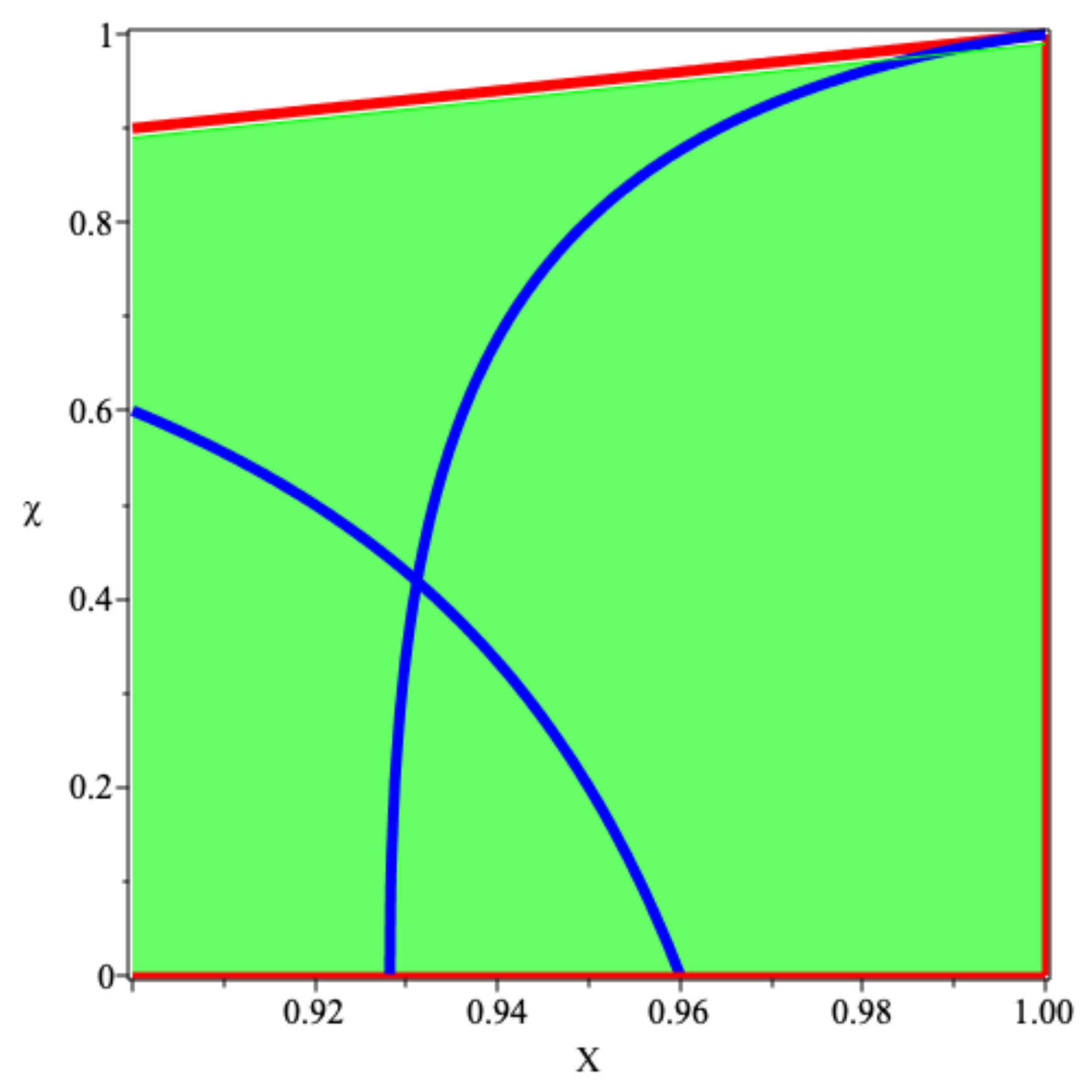}
\caption{Subset of the physical region $0 \leq \chi \leq \Chi \leq 1$ where both $w<1$ and $F>1/4$. 
The condition $w<1$ (the DEC) is satisfied below the monotone decreasing curved line. The condition $F>1/4$ (violating the strong maximum force conjecture) is satisfied below the monotone increasing curved line.
There is a region between  $\Chi = 4\sqrt{3}-6 = 0.928203232...$ and $\Chi = {24\over25} = 0.96$ where for a finite range of $\chi$ it is possible to satisfy the DEC while violating the strong maximum force conjecture.
By considering the region above the two curved lines, it is also possible to violate the DEC while satisfying  the (strong) maximum force conjecture. 
Similarly to the left of the two curved lines both DEC and maximum force conjectures are satisfied, while to the right of the two curved lines both DEC and maximum force conjectures are violated.
}
\label{F-superimpose}
\end{center}
\end{figure}

By superimposing the parts of the physical regime $0 \leq \chi \leq \Chi \leq 1$ where both $w<1$ and $F>1/4$ one easily sees that it is possible to 
violate the (strong) maximum force conjecture without violating the DEC.  See figure \ref{F-superimpose}. 
It is also possible to violate the DEC while satisfying  the (strong) maximum force conjecture. 
For this to occur one must be ``close'' to black hole formation $\chi\lesssim\Chi\lesssim1$. 

\clearpage

\section{Weak-gravity regime}\label{S:weak-gravity}
\enlargethispage{40pt}
In the weak-gravity regime, (where both $\Chi = 2M/a\ll 1$ and $\chi=2m/a\ll1$), we have
\begin{equation}
\sigma \approx {1\over 4\pi } {M-m\over a^2}  \geq 0,
\end{equation}
and
\begin{equation} 
\Pi \approx {1\over 16\pi } {M^2-m^2\over a^3}  \geq 0. 
\end{equation}

So in the weak-gravity regime one has the physically plausible result that 
\begin{equation}
M\approx m+ 4\pi \sigma \,a^2,
\end{equation}
while the surface stress is 
\begin{equation}
\Pi \approx  {1\over 4} \; {(M+m)\over a} \; \sigma \;\ll\;  \sigma. 
\end{equation}
That is
\begin{equation}
w = {\Pi\over\sigma} \approx {1\over 4} \; {(M+m)\over a}
={\Chi+\chi\over 8}
 \; \;\ll\; 1. 
\end{equation}
So all the classical energy conditions, including the DEC, are automatically satisfied in the weak-gravity regime. 

Finally note that in the weak gravity regime the force between two hemispheres is 
\begin{equation}
F \approx {1\over 8 } \; {M^2-m^2\over a^2} 
= 
{1\over2} (\Chi^2-\chi^2)  \ll 1. 
\end{equation}
However, we have seen above that we have much more general and precise results available, fully applicable to the highly relativistic strong-gravity regime.

\section{Low-mass thin-shell configurations}\label{S:low-mass}
\newcommand*{\wasyfamily}{\fontencoding{U}\fontfamily{wasy}\selectfont}
\newcommand*{\astrosun}{{\odot}}
\newcommand*{\mercury}{{\text{\wasyfamily\char39}}}
\newcommand*{\venus}{{\text{\wasyfamily\char25}}}
\newcommand*{\earth}{{\oplus}}
\newcommand*{\mars}{{\text{\wasyfamily\char26}}}
\newcommand*{\jupiter}{{\text{\wasyfamily\char88}}}
\newcommand*{\saturn}{{\text{\wasyfamily\char89}}}
\newcommand*{\uranus}{{\text{\wasyfamily\char90}}}
\newcommand*{\neptune}{{\text{\wasyfamily\char91}}}
\newcommand*{\pluto}{{\text{\wasyfamily\char92}}}

Another interesting approximation is that of a low-mass thin-shell, where
\begin{equation}
M - m \ll m.
\end{equation}
We can then approximate
\begin{equation}
\sigma(m,M;a) = {M-m\over 4\pi a^2 \sqrt{1-2M/a} }+\O([M-m]^2),
\end{equation}
while
\begin{equation}
\Pi(m,M;a) ={m(M-m)\over  8\pi a^3 \sqrt{1-2M/a}} +\O([M-m]^2).
\end{equation}

\clearpage
Partially eliminating $M-m$ between these two equations we see
\begin{equation}
\Pi(m,M;a) = {1\over 4}\;  {2m\over a} \; \sigma(m,M;a)  + \O([M-m]^2).
\end{equation}

Trivially,
\begin{equation}
w(m,M;a) ={m/a\over  2(1-2m/a)} + \O([M-m]),
\end{equation}
and 
\begin{equation}
F(m,M;a) ={m(M-m)\over  a^2 \sqrt{1-2M/a}} +\O([M-m]^2),
\end{equation}
while with a little more work we can obtain
\begin{equation}
w(m,M;a) ={m/a\over  2(1-2m/a)} 
+{(M-m)\over 4 a(1-2m/a)^2} +
\O([M-m]^2).
\end{equation}
Thus even for a low mass shell, $M - m \ll m$, by driving $\chi \to 1$, one can still easily arrange violations of the DEC and the original form of the maximum force conjecture.

To calibrate the approximate size of these quantities note that in geometrodynamic units the mass of our Sun is $
M_\astrosun \approx 1.5 \;\hbox{km}$, while the mass of Jupiter is $M_\jupiter\approx 1.1 \;\hbox{m}$.  The astronomical unit is $1 \;\hbox{AU} \approx 1.5\times 10^8 \;\hbox{km}$.  Then disassembling Jupiter to build a mega-sphere with radius 1 AU leads to the dimensionless estimates 
\begin{equation}
w \approx 5 \times 10^{-9};\qquad 
F \approx 7.333 \times 10^{-20}.
\end{equation}
That is, solar system scale Dyson mega-spheres would be well into the Newtonian regime. 
For general relativistic effects to be appreciable one must be ``close'' to black hole formation $\chi\lesssim\Chi\lesssim1$. 

Furthermore in physical units we have 
$M_\jupiter\approx 1.898 \times 10^{27}\;\hbox{kg}$, 
so that  for the surface energy density
\begin{equation}
\sigma\approx {M_\jupiter\over (4\pi \,1 \, {\hbox{AU}}^2)} 
\approx 7003 \;\hbox{kg/m}^2. 
\end{equation}
Locally, this is actually human scale --- since the density of iron is $\approx 7800 \;\hbox{kg/m}^3$, this 
would correspond to a Dyson mega-sphere a little under 1 metre thick. Globally speaking however, this is well beyond human technology, and would at the very least require the intervention of a  Kardashev type II civilization~\cite{Kardashev:1,Kardashev:2}.

\clearpage
\section{Possible generalizations}\label{S:geeralize}

\subsection{Matrioshka configurations}\label{S:Matrioshka}

It would in principle be possible to extend the current  discussion to deal with the more general Matrioshka configurations (multiply nested thin-shell Dyson mega-spheres). Such configurations have been mooted in the context of Matrioshka brains~\cite{Matrioshka}. 
No really new matters of principle would be involved, but the algebra would become considerably more complex. We put such issues aside for now.

\subsection{Thick-shell Dyson mega-spheres}\label{S:thick}

In contrast potential consideration of thick-shell Dyson mega-spheres would require a completely different set of techniques --- one would need to study the \emph{anisotropic} Tolman--Oppenheimer--Volkoff (TOV) equation, paying careful attention to suitable boundary conditions and equations of state. We put such issues aside for now. 

\subsection{Adding rotation}\label{S:rotation}

Adding rotation to these Dyson mega-sphere configurations would likely be extremely tedious, if not outright impractical. Even if the angular momentum of the central object and Dyson mega-sphere are parallel, the required calculations would be daunting. One could either try to work with the full Kerr geometry~\cite{Kerr,Kerr-Texas,kerr-newman,kerr-intro,kerr-book,kerr-book-2}, or with the Lense--Thirring slow rotation approximation~\cite{Lense-Thirring,Pfister,PGLT1,PGLT2,PGLT3,PGLT4}. We put such issues aside for now.

\section{Conclusions}\label{S:discussion}

Arbitrarily advanced alien civilizations are the \emph{sine qua non} of much speculative physics --- developed with a view to testing the ultimate limitations of putative technological advancement. 
Notable examples of speculative physics are Dyson mega-structures, (specifically the Dyson mega-spheres of the current article), Lorentzian wormholes, warp drives, and tractor/pressor/stressor beams. 
A common theme arising in such speculative physics is the violation of some or all of the classical energy conditions, 
and/or violations of the original strong form of the maximum force conjecture. 
Such violations are not the absolute prohibition they were once thought to be, but they are certainly invitations to think deeply about the underlying physics. 

\clearpage
The Dyson mega-structures in particular, are in principle accessible to astronomical observational tests. In the current article we have analyzed the simple spherically symmetric thin-shell Dyson mega-sphere, and have done so in a fully general relativistic manner using the Israel--Lanczos--Sen thin-shell formalism. 
We have seen that there are regions in parameter-space where either, both, or neither of the DEC and strong maximum quasi-local force conjecture are violated. 

\section*{Acknowledgements}

TB was indirectly supported by the Marsden Fund, 
\emph{via} a grant administered by the Royal Society of New Zealand.\\
AS was supported by a Victoria University of Wellington PhD scholarship, and was also indirectly supported by the Marsden Fund, \emph{via} a grant administered by the Royal Society of New Zealand.\\
MV was directly supported by the Marsden Fund, \emph{via} a grant administered by the Royal Society of New Zealand.\\

\hrule\hrule\hrule


\end{document}